\documentclass[aps,prb,twocolumn,showpacs]{revtex4}
\usepackage{pstricks}
\usepackage{graphicx}
\usepackage{epsfig}
\usepackage{multirow}

\begin{document}

\title{Spin and exchange coupling for Ti embedded in a surface 
dipolar network} 

\author{Raghani Pushpa*$^{1,2,4}$}
\author{Jesus Cruz$^{3,2}$} 
\author{Barbara Jones$^2$}
\affiliation{$^1$Center for Probing the Nanoscale, Stanford University, Stanford,
California, USA}  
\affiliation{$^2$IBM, Almaden Research Center, San Jose, CA, USA}
\affiliation{$^3$Georgetown University, Washington, D.C., USA} 
\affiliation{$^4$Boise State University, Boise, Idaho, USA} 

\date{\today}

\begin{abstract}
We have studied the spin and exchange coupling of 
Ti atoms on a Cu$_2$N/Cu(100) surface using
density functional theory. We find 
that individual Ti have a spin of 1.0 (i.e., 2 Bohr Magneton) on the Cu$_2$N/Cu(100) 
surface instead of spin-1/2
as found by Scanning Tunneling Microscope. We suggest an explanation for this
difference, a two-stage Kondo effect, which can be verified by experiments. 
By calculating the exchange coupling for Ti dimers on the Cu$_2$N/Cu(100) surface, 
we find that the exchange coupling across a `void' of 3.6\AA\ is antiferromagnetic, 
whereas indirect (superexchange) coupling through a 
N atom is ferromagnetic. 
For a square lattice of Ti on Cu$_2$N/Cu(100), we find a novel spin striped phase. 

\end{abstract}
\pacs{
71.15.Mb, 
71.70.Gm, 
71.15.Nc, 
68.55.-a, 
}

\maketitle

\section{\label{introd}Introduction}
Atomic-scale magnetic structures on surfaces \cite{Cyrus-exchange} 
are of current interest for several reasons. Primarily, they display
intriguing physical properties in their own right. Magnetic atoms
on surfaces, simple or complex, can display large magnetocrystalline
anisotropy which differs from the bulk. The spin can be large,
or quenched by electronic effects such as the Kondo effect.
And coupling between spins can be via direct overlap, RKKY or 
superexchange. Secondarily, these systems are compelling because 
of their parallels with other nano-scale systems -- quantum dots, 
magnetic multilayers, magnetic impurities in thin films, to name
just a few. Finally, there are the possible applications to nano-scale
magnetic bits and future magnetic devices.
A large net spin and magnetic anisotropy are required for
atomic-scale magnetic structures to be used as practical 
nano-scale magnetic bits.
A possible way to obtain
a large magnetic moment is through a ferromagnetic 
coupling between transition metal atoms. However, ferromagnetic 
coupling is rare in transition metal complexes, \cite{Kahn}
that is, when the transition metal atom is bonded to a nonmetallic
atom. We describe below our studies of such a system.

In a Scanning Tunneling Microscope (STM) measurement 
\cite{otte} of a Ti atom
placed on a Cu$_2$N/Cu(100) surface, it was found that the Ti exhibits very 
different magnetic properties than in gas phase.
In the following work, we use density functional theory (DFT) to study
the atomic spin of a single Ti atom,
and exchange coupling of a 
dimer of Ti atoms, placed on a single layer of Cu$_2$N on a Cu(100) surface.
The Cu$_2$N layer is used as an insulating 
layer to isolate the spin of a magnetic adatom from the metal electrons
of the Cu(100) surface \cite{Cyrus-exchange,Cyrus}. 
Hereafter, the Cu$_2$N/Cu(100) surface will be referred to as the CuN surface.  
We study exchange coupling between Ti atoms in two different 
environments: (i) a square lattice of Ti on the CuN surface and
(ii) a dimer of Ti atoms deposited on the CuN surface.


\section{DFT calculations}

We use spin-polarized DFT, as implemented in Quantum-ESPRESSO \cite{QE},
within a
pseudopotential formalism using a plane wave basis with 
a cut-off of
30 Ry. A higher cut-off of 240 Ry was used for the
augmentation charges introduced by the ultrasoft pseudopotential
\cite{uspp}. 
We use the generalized gradient approximation
(GGA) for the exchange correlation interaction with the functional
proposed by Perdew, Burke and Ernzerhof. \cite{PBE} An
on-site Coulomb interaction (U) for Ti was employed, with $U$ taken 
to be 4.7 eV as calculated using a constraint-GGA method 
\cite{Madsen,CYL}.
To improve the convergence, a gaussian smearing of width
0.01 Ry was adopted. Brillouin zone integrations for the ($1\times
1$) surface cell of Cu(100) were
carried out using a ($16\times 16\times 1$) mesh of $k$-points.

We obtained the bulk lattice parameter for Cu as 3.67 \AA ,
which compares
well with the experimental value of 3.61 \AA \cite{Ashcroft}.
To simulate the CuN surface, we use a symmetric slab of three
to five atomic layers of Cu, with a layer of CuN above and
below. Periodic images of the slab were 
separated by a vacuum of 15 \AA\ along the z $\left [ 100 \right ]$ direction.

\section{CuN surface}
\begin{figure}
\includegraphics[width=84mm]{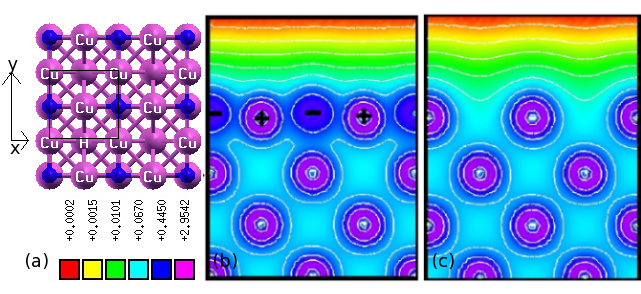}
\caption{(a) Top view of the CuN on Cu(100) surface; 
small and big spheres represent
N and Cu atoms respectively. Big spheres with symbol ``Cu" are 
the top layer Cu atoms and those without the symbol are the
second layer. 
The black square indicates the c($1\times1$)
unit cell of the CuN surface. Along the x-axis, surface 
Cu atoms are separated by a hollow site (it is marked by `H')
and along the y-axis, they 
are separated by a N atom.
Side views of charge density produced by a 
cut along the (b) N-axis and (c) hollow-axis.}
\label{CuN-strct-Charge}
\end{figure}

The top view of the CuN 
surface is shown in Fig. \ref{CuN-strct-Charge}a.
The unit cell of the CuN surface
(shown by the black square) consists of two Cu atoms and one N atom.
We find that the N atoms are 0.18 \AA\ 
above the top Cu atoms in a fully relaxed structure. The distance 
between the first and second layers of Cu is 1.97 \AA\ which compares 
well with the all-electron result of 1.91 \AA\ \cite{Maria}.  
We find that each Cu atom in the 
surface unit cell loses 0.7 electrons to the N 
(presumably due to the electronegative nature of N). 
Thus, Cu and N atoms form a square network of dipoles on the surface, 
rendering a (nominally) insulating character to the surface. 
As shown in Fig.\ref{CuN-strct-Charge}a, 
along the x-axis, two surface Cu atoms
are separated by a hollow site and along the y-axis, they are separated
by a N-atom. These directions will be referred as hollow-axis and
N-axis, respectively.
The charge 
density of the CuN surface 
along the N- and hollow-axis is shown in Fig.\ref{CuN-strct-Charge}(b) and (c)
respectively. 
Notice that along the N-axis, charge gets accumulated in the 
top layer.

\section{Spin of Ti on CuN surface}
To calculate the spin of Ti on the CuN surface, a Ti atom is deposited
on top of a Cu atom, following the STM experiments \cite{otte}. At
one monolayer coverage of Ti, the ($1\times1$) surface  
unit cell consists  
of one N atom, one Ti atom and two Cu atoms
as shown in Fig.\ref{CuN-strct-Charge}a.  
A constrained-GGA \cite{Madsen} calculation \cite{CYL}
yields $U=4.7$ eV for Ti in this configuration.
However, the value of $U$ for Ti would
presumably be different at lower coverages of Ti.  
In order to understand the effect of U on the spin of Ti, we do calculations
for a range of values of $U$ for the ($1\times1$) unit cell.
The results for distances between Ti and its nearest neighbor atoms
in the surface, angle subtended by Ti-N-Ti, and 
the spin of Ti are shown in Table \ref{Structure}.
We find that Ti-N and Ti-Cu distances increase as $U$ increases,
with the net effect of a rising Ti and decreasing (becoming sharper) Ti-N-Ti angle.
Most importantly, as U increases the spin of Ti approaches that of 
the gas phase value of 1. 

As an added complexity, we find that the
initial magnetization of Ti affects the final calculated
ground state, 
indicating a complex energy minimization landscape. Hence 
we try several initial magnetizations and take the final state corresponding 
to the lowest energy.
For U=4.7, we show the data corresponding to two such optimized 
structures (S-I and S-II) obtained by varying the initial magnetization
(see Table \ref{Structure}).
The optimized 
structure corresponding to S-I is the 
lowest energy structure (S-I is lower in energy than S-II by 0.3 eV), 
showing the spin of Ti to be 0.75, 
indicating possible mixed valent behavior for a monolayer of Ti.

\begin{table*} 
\begin{tabular}{|c|c|c|c|c|c|c|} 
\hline
System & Cell  & $U$ & $d_{Ti-N}$ & $d_{Ti-Cu}$ & $A_{Ti-N-Ti}$ & $S$  \\ 
\cline{1-7}
\multirow{8}{*}& \multirow{5}{*}{$ 1 \times 1 $} & 0.0 & 1.91 & 2.49 & 148.6 & 0.0  \\
Single & & 3.0 & 1.95 & 2.54 & 141.6 & 0.6   \\
Ti & & \bf{4.7 (S-I)} & \bf{1.99} & \bf{2.56} & \bf{135.6} & \bf{0.75} \\
Atom & & 4.7 (S-II) & 2.27 & 2.68 & 108.4 & 1.0  \\ 
& & 6.0 & 2.33 & 2.72 & 104.2 & 1.0  \\ 
\cline{2-7}
 & {$ 2 \times 2 $}  & 4.7 & 2.07 & 2.58 & - & 1.0  \\
 & {$ 2 \times 3 $}  & 3.0 & 1.98 & 2.52 & - & 1.0  \\
 & {$ 3 \times 3 $}  & 6.0 & 2.13 & 2.66 & - & 1.0 \\ \hline
\multirow{2}{*}{Ti Dimer} & N-axis  & 4.7 & 2.04 & 2.65 & 142.9 & 1.0 \\ 
& H-axis  & 4.7 & 2.09 & 2.56 & - & 1.0  \\ \hline

\end{tabular}
\caption{The Ti-N bond length ($d_{Ti-N}$), the Ti-Cu bond length ($d_{Ti-Cu}$), 
the Ti-N-Ti angle ($A_{Ti-N-Ti}$) and the spin $S$ of the Ti atom 
on the CuN surface.
The top panel shows these results as a function of Hubbard $U$ (in eV) on Ti, 
for a ($1\times1$) unit cell.
The middle panel shows these results for a single Ti atom in ($2\times2$), ($2\times3$), and 
($3\times3$) unit cells. The bottom panel shows these results for a dimer
of Ti adsorbed along the N- and hollow-axis, respectively.
All the bond lengths are given in Angstroms.}
\label{Structure}
\end{table*}

Next, we calculate the spin of Ti in ($2\times2$), ($2\times3$),
and ($3\times3$) unit cells, i.e., at coverages of 1/4, 
1/6, and 1/9 ML respectively. This
data is shown in the middle panel of Table \ref{Structure}.
Comparable to the ($1\times1$) unit cell, the distance 
of Ti from the Cu below in ($2\times2$) is 2.58 \AA. There is also a similar
trend of Ti rising higher 
above the surface than N, by an amount increasing with 
increasing U. 
Most importantly, it was found that in all the three cases
of larger unit cells, 
the spin on Ti is 1. At this point, we note a
discrepancy with STM\cite{otte} experiments, which see a spin-1/2 Kondo effect.
We postulate a resolution of the issue with a two-stage Kondo effect, 
in which the spin is first compensated from spin-1 to spin-1/2 at a 
higher temperature, and then seen as a spin-1/2 Kondo effect at 
the experimentally observed temperature.

\section{Exchange coupling}
To calculate the exchange coupling,
we assume a Heisenberg interaction
($H=J\bf{S_1.S_2}$), and can relate the value of J to the energy
difference between ferromagnetic and (Ising) antiferromagnetic
configurations: 

\begin{equation}
2S^2J = E_{\uparrow\uparrow}-E_{\uparrow\downarrow} \equiv \Delta E
\label{Jeqn}
\end{equation}

Here, $S$ is the magnitude of spin,
and $J$ is the exchange 
coupling. $E_{\uparrow\uparrow}$ and $E_{\uparrow\downarrow}$ are 
the total energies calculated from DFT when the spins on the magnetic
atoms point along the same direction and in 
opposite directions respectively. Note that Eq. \ref{Jeqn} holds for all
values of quantum spin. The relationship with J is valid for each
$S^z$ always at full maximal or minimal value 
(Ising antiferromagnet; colinear spins);
for the energy difference with a full quantum antiferromagnetic state, 
the term $2S^2J$ would become $(2S+1)SJ$. 
In this paper, we will mainly concentrate on the energy difference 
between ferromagnetic (aligned) and antiferromagnetic (antialigned)
configurations, rather than on the value of J. We calculate the exchange 
coupling for Ti lattices (1ML coverage of Ti) and for two Ti atoms 
placed on the CuN surface in a large unit cell.

\subsection{Lattice of Ti atoms on CuN surface}
\begin{figure}
\includegraphics[width=84mm]{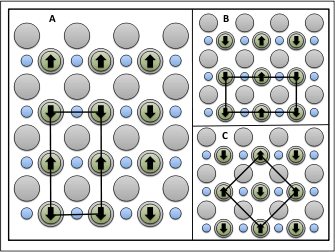}
\caption{Schematic diagrams showing spin configurations in
Ti lattices. In configuration (a), spins are aligned along
the N-axis and antialigned along the hollow-axis; in (b) spins are antialigned
along the N-axis and aligned along the hollow-axis. Configuration (c) is
a checkerboard configuration with spins antialigned along
both the N- and hollow-axis.}
\label{Hollow}
\end{figure}

At one monolayer coverage of Ti on CuN surface, Ti forms a square ``lattice''
on the surface. In this case, the energy of the ferromagnetic configuration ($E_{FM}$)
is the total energy 
of the ($1\times1$) unit cell since it contains only one Ti atom. 
However, to obtain $E_{\uparrow\downarrow}$, we design three different 
configurations with ($1\times2$), ($2\times1$), and ($\sqrt 2\times\sqrt 2$)
unit cells as shown in Fig. \ref{Hollow}(a), (b), and (c) respectively. Arrow signs in the figure indicate 
relative direction of spins on Ti atoms. Total energies of the three configurations
will be referred to as $E_H$, $E_N$, and $E_C$ respectively. Notice
that the unit cell size in all three configurations is twice that of the
ferromagnetic configuration.  
Subtracting the total energies of configurations (a), (b), and (c) 
from two times the energy of the ferromagnetic 
configuration ($2\times E_{FM}$)
will give the exchange coupling of 
Ti atoms along the hollow-axis, along the N-axis, and in the checkerboard 
configuration,
respectively; assuming that there are only nearest neighbor interactions.

Our results for exchange coupling are summarized in 
Table \ref{Exchange-Coupling}.
For the lowest energy structure S-I, we find that the exchange coupling along the N-axis is 
unexpectedly ferromagnetic, i.e., the total energy $E_{FM}$
is lower than $E_N$ by 16.1 meV.
However, the exchange coupling across a hollow is antiferromagnetic,
i.e., the total energy $E_{H}$
is lower than $E_{FM}$ by 106.8 meV.
Thus, the antiferromagnetic coupling along the hollow-axis is much 
stronger than the ferromagnetic coupling along the N-axis.
The checkerboard pattern (Ising antiferromagnet) is more
favored over a pure ferromagnetic state with 
$\Delta E$ being 77.8 eV; however, it is 
less favorable than the hollow-axis antiferromagnetism, 
presumably due to the energy disadvantage of antiferromagnetic coupling
along the N-axis. The overall order, from lowest to highest energy,
is $E_H < E_C < E_{FM} < E_N$.
Configuration Fig. \ref{Hollow}a is the ground state
and we term it a spin striped state.
These ferromagnetic stripes should be observable in large 
enough lattices.

\begin{figure}
\includegraphics[width=84mm]{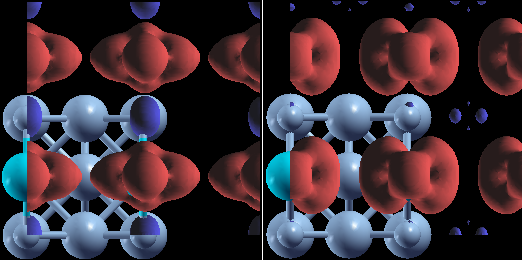}
\caption{Spin density plot for ($1\times1$) unit cell for lower energy
(a) and higher (b) configurations. Different colors (shadings in 
black and white version) correspond to 
opposite spins. The Nitrogen atoms appear as small spheres with
opposite polarization between the Ti.}
\label{Spin-1x1}
\end{figure}

In order to understand how structure plays a role in the exchange coupling,
we also calculate spin exchange for the structure S-II 
(Table \ref{Exchange-Coupling}). We notice that the 
exchange coupling for the structure S-II is much lower than that 
of S-I. This could possibly be due to lower interaction of 
Ti with the surface (See Table-\ref{Structure}, Ti-Cu and 
Ti-N distances are longer in S-II than those in S-I).
Spin density plots for the two structures in the ferromagnetic
state
are shown in Fig.\ref{Spin-1x1}. Notice that the spin density
gets stretched out along the hollow-axis for S-I. 
Also, the N atoms get spin polarized for S-I more than
for S-II. This shows that in S-I,
there are stronger interactions. The net result is that this
structure has the lowest total energy.

\begin{table}
\begin{tabular}{|c|c|c|c|}
\hline
System & Structure & $ E_{FM} - E_{N}$ & $ E_{FM} - E_{H}$ \\
\hline
\multirow{2}{*}{Lattice $1 \times 1 $ }& S-I & -16.1 & 106.8 \\ 
& S-II & 6.2 & 13.9 \\ 
\hline
\ Dimer $2\times3$ & - & -16.5 &  143.9 \\ 
\hline
\end{tabular}
\caption{The energy differences $\Delta E$ along the N-axis ($E_{FM}-E_N$) and 
the hollow-axis ($E_{FM}-E_H$)
for a lattice of Ti in a ($1\times1$) unit cell,  and a dimer of Ti in a ($2\times3$) unit cell.}
\label{Exchange-Coupling}
\end{table}

\subsection{Dimer of Ti atoms on CuN surface}
We have drawn conclusions so far about Ti-Ti coupling
based on the calculations in lattices, where the situation is 
more complicated because one not only has the nearest neighbor
(NN) interactions but also has next NN (NNN) interactions 
and so on. To simulate a Ti-Ti dimer on the surface
we use 
a larger unit cell of ($2\times3$) with two and three lattice units along
the hollow- and N-axis respectively.
Interestingly, we find ferromagnetic coupling along the N-axis
and antiferromagnetic along the hollow-axis, the same ground 
states as for the  
case of ($1\times1$) lattices. Along the N-axis, the energy
difference ($\Delta E$) is -16.5 meV compared to -16.1 meV for 
the ($1\times1$) case. Along the hollow-axis the energy difference
is 143.9 meV compared to 106.8 meV for the ($1\times1$) case. 
Thus, a Ti lattice and a dimer show a similar trend and strength of coupling.
It confirms our assumption of primarily nearest neighbor interactions in 
a Ti lattice on the CuN surface.
Notice that the distance between Ti and the Cu atom below it is 
2.56 \AA\ for both the ($1\times1$) case, and the ($2\times3$) case 
for coupling along the hollow-axis.
However, when the dimer is placed along the N-axis the 
distance between Ti and the Cu below it increases slightly to 2.65 \AA.
The Ti-N-Ti angle is 135.6 degrees for the ($1\times1$) case which is
close to
142.9 degrees for the ($2\times3$) case.

\begin{figure}
\includegraphics[width=60mm]{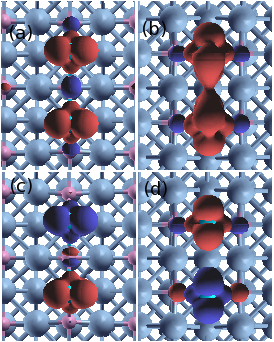}
\caption{Spin density ($\rho_{\uparrow}-\rho_{\downarrow}$)
plot for ferromagnetic (top pannel: (a) and (b))
and antiferromagnetic (bottom pannel: (c) and (d)) configuration of 
a Ti dimer along 
the (a) N-axis and (b) hollow-axis. The most energetically favorable
configurations are ferromagnetic across a N (top left) and 
antiferromagnetic across a void (bottom right).}
\label{Spin-dimer}
\end{figure}

In Fig. \ref{Spin-dimer}, we plot the spin density for the Ti dimer along
the N-axis (Fig.\ref{Spin-dimer}a) and the hollow-axis 
 (Fig.\ref{Spin-dimer}b).
A significant amount of induced 
spin-polarization around the N atom can be seen from the figure.  
Ferromagnetic coupling
between Ti atoms along the N-axis is established by having an 
opposite spin N atom both between the Ti atoms and at 
opposite ends.
For anti-aligned spin configuration along the N-axis, the N atom
becomes a single-atom antiferromagnet with a net spin of zero.
Along the hollow-axis, when spin on both the Ti atoms is 
aligned, a dramatic
anisotropy in the spin polarization of the Ti develops,
with a direct overlap established over the hollow site
(Fig.\ref{Spin-dimer}b). 
The stretching of the Ti bonds in this case
case is striking, and suggestive that higher symmetry considerations
may be coming into play.
However, when spins are antialigned, no such elongation of 
spin polarization occurs. In both the cases, N atoms 
on the sides of the two Ti atoms develop a spin polarization 
opposite to that of the Ti.

The primary sources of exchange coupling 
between the Ti atoms are superexchange \cite{Bencini}, RKKY \cite{RKKY},
and direct overlap/direct exchange \cite{Ashcroft}. The coupling between the adatoms 
can be direct, if the wave functions should
overlap, or RKKY, if the influence of the Cu in the layers below
is strong enough. Along the N-axis, the center N atom
becomes a
natural source for a superexchange coupling between Ti atoms,
ruling out RKKY which 
would need to 
take an indirect route under the N atom, a much longer route 
than directly across the N for superexchange.
Along the
hollow-axis, however, there is no convenient single atom to 
hop across for superexchange, rather the sea of conduction electrons
from the underlying and intervening Cu. (Unless one is to consider
superexchange via the second-layer Cu, an unlikely candidate.)
In this case RKKY and direct
overlap become more likely. Indeed for an aligned spin configuration,
we observe a direct overlap forming, as discussed above.
However, the lowest energy state for coupling along the hollow-axis
is antiferromagnetic, and we conclude that in this case it is likely 
due to RKKY coupling. This could be tested experimentally by varying 
the Ti-Ti distance and measuring the exchange coupling; however, only certain
discrete lattice positions would be possible.

\section{Conclusions}
We find that a Ti atom has spin-1 on the CuN surface instead 
of spin-1/2 as found in the experiments \cite{Cyrus}. As a possible
explanation, we propose that there should be a two-stage Kondo effect
for this system. At high temperatures, Ti starts with spin-1,
which then undergoes a transition to spin-1/2 with a first-stage
Kondo effect. At a lower temperature the second stage Kondo effect
brings it down from spin-1/2 to a spin-zero, and it is this spin-1/2 
Kondo effect that is seen in the low-temperature STM. This
prediction can be tested by measuring the local density of states
at higher temperatures than 0.5 K.

We find a ferromagnetic coupling along the N-axis and antiferromagnetic
along the hollow-axis, for both the lattice and dimer of Ti on the CuN
surface.
Ti lattice and dimer have a similar trend as well as strength
of coupling. This 
indicates that interactions between Ti atoms in the lattice 
configuration are local;
and a marked spin striped phase is found as the 
ground state of the lattice. We find a ferromagnetic coupling along the N-axis
due to superexchange, with secondary contributions from direct exchange.
We also postulate that the antiferromagnetic coupling along the hollow-axis is 
primarily due to RKKY interactions, with a smaller direct exchange component.

\end{document}